\documentclass[12pt]{article}
\usepackage{amsthm,latexsym,multibox,amssymb,amsfonts,graphicx}

\def\a{\alpha}
\def\b{\beta}
\def\g{\gamma}
\def\d{\delta}

\def\h{\eta}
\def\l{\lambda}

\def\m{\mu}
\def\n{\nu}
\def\r{\rho}
\def\o{\omega}
\def\s{\sigma}
\def\S{\Sigma}
\def\t{\tau}
 
\def\x{\xi}

\def\ve{\varepsilon}

\def\pa{\partial}

\def\be{\begin{equation}}
\def\ee{\end{equation}}
\def\bqn{\begin{eqnarray}}
\def\eqn{\end{eqnarray}}

\def\cs{{\cal S}}

\begin{document}

\begin{titlepage}
\begin{flushright}
ULB-TH-03/21 \\ 
\end{flushright}
\vskip 2.5cm

\begin{centering}

{\large {\bf A note on spin-$s$ duality}}

\vspace{2cm}

Nicolas Boulanger$^{a,\sharp}$, Sandrine Cnockaert$^{a,1}$ and
Marc Henneaux$^{a,b}$ \\
\vspace{.7cm} {\small $^a$ Physique Th\'eorique et Math\'ematique,
Universit\'e Libre
de Bruxelles, C.P. 231, B-1050, Bruxelles, Belgium \\ 
\vspace{.2cm} $^b$ Centro de Estudios Cient\'{\i}ficos, Casilla
1469, Valdivia, Chile }

\vspace{2cm}

\end{centering}

\begin{abstract}
Duality is investigated for higher spin ($s \geq 2$), free,
massless, bosonic gauge fields. We show how the dual formulations 
can be derived from a common ``parent", first-order action. This
goes beyond most of the previous treatments where higher-spin
duality was investigated at the level of the equations of motion
only. In $D=4$ spacetime dimensions, the dual theories turn out to
be described by the same Pauli-Fierz ($s=2$) or Fronsdal ($s \geq
3$) action (as it is the case for spin $1$). In the particular
$s=2$ $D=5$ case, the Pauli-Fierz action and the Curtright action
are shown to be related through duality. A crucial ingredient of 
the analysis is given by the first-order, gauge-like,
reformulation of higher spin theories due to Vasiliev.
\end{abstract}

\vspace{5.5cm} 
\noindent \footnotesize{$\sharp$ ``Chercheur F.R.I.A.'' (Belgium)}
\\ \noindent 
\footnotesize{$^1$ Aspirante du Fonds National de 
la Recherche Scientifique (Belgium)}

\vfill
\end{titlepage}
%
%********************************
\section{Introduction} 
\setcounter{equation}{0}
\setcounter{theorem}{0}
\setcounter{lemma}{0}
%********************************
%
Duality for higher spin massless gauge fields has been the focus
of a great interest recently 
\cite{Nieto:1999pn,CasUrr0,Hull1,BekBoul,HulldM1,Bekaert:2003az,CasUrr1}. 
In most of this recent work, duality is studied at the level of the
equations of motion only (notable exceptions being
\cite{Nieto:1999pn,CasUrr0,CasUrr1}, which deal with the spin-2 case in 4
spacetime dimensions). One may wonder whether a stronger form of
duality exists, for all spins and in all spacetime dimensions, 
which would relate the corresponding actions. A familiar example
in which duality goes beyond mere on-shell equivalence is given by
a set of a free $p$-form gauge field and a free $(D-p-2)$-form
gauge field in $D$ spacetime dimensions. The easiest way to
establish the equivalence of the two theories in that case is to
start from a first order ``mother" action involving simultaneously
the $p$-form gauge field $A_{\mu_1 \cdots \mu_p}$ and the field 
strength $H_{\mu_1 \cdots \mu_{D-p-1}}$ of the $(D-p-2)$-form
$B_{\mu_1 \cdots \mu_{D-p-2}}$ treated as independent variables
\begin{equation}
S[A,H] \sim \int dA \wedge H - \frac{1}{2}H \wedge \! ^*H
\label{action0}
\end{equation} 
The field $H$ is an auxiliary field that can be eliminated through
its own equation of motion, which reads $H = \! ^* dA$. Inserting
this relation in the action (\ref{action0}) yields the familiar
second-order Maxwell action $\sim \int dA \wedge \! ^* dA$ for
$A$. Conversely, one may view $A$ as a Lagrange multiplier for the
constraint $dH = 0$, which implies $H = dB$. Solving for the
constraint inside (\ref{action0}) yields the familiar second-order
action $\sim \int dB \wedge \! ^* dB$ for $B$.

Following Fradkin and Tseytlin \cite{Fradkin:1984ai}, we shall
reserve the terminology ``dual theories" for theories that can be
related through a ``parent action", referring to ``pseudo-duality"
for situations when there is only on-shell equivalence. The parent
action may not be unique. In the above example, there is another,
``father" action in which the roles of $A$ and $B$ are
interchanged ($B$ and $F$ are the independent variables, with $S 
\sim \int dB \wedge F - \frac{1}{2} F \wedge \! ^*F$ and $F = dA$
on-shell). That the action of dual theories can be related through
the above transformations is important for establishing
equivalence of the (local) ultraviolet quantum properties of the
theories, since these transformations can formally be implemented 
in the path integral \cite{Fradkin:1984ai}.

Recently, dual formulations of massless spin-2 fields have
attracted interest in connection with their possible role in
uncovering the hidden symmetries of gravitational theories
\cite{CJLP,OP1998,West,H-L1,Damour,Henry-Labordere:2002xh,EHTW}. In these
formulations, the massless spin-2 field is described by a tensor 
gauge field with mixed Young symmetry type. The corresponding
Young diagram has two columns, one with $D-3$ boxes and the other
with one box. The action and gauge symmetries of these dual
gravitational formulations have been given in the free case by
Curtright \cite{Curtright}. However, the connection with the more
familiar Pauli-Fierz formulation \cite{PaFi} was not clear and
direct attempt to prove equivalence met problems with trace 
conditions on some fields. The difficulty that makes the spin-$1$
treatment not straightforwardly generalizable is that the
higher-spin ($s \geq 2$) gauge Lagrangians are not expressed in
terms of strictly gauge-invariant objects, so that gauge
invariance is a more subtle guide. One of the purposes of this
note is to show explicitly that the Curtright action and the
Pauli-Fierz action both come from the same parent action and are 
thus dual in the Fradkin-Tseytlin sense. The analysis is carried
out in any number of spacetime dimensions and has the useful
property, in the self-dual dimension four, that both the original
and the dual formulations are described by the same Pauli-Fierz
Lagrangian and variables. 

We then extend the analysis to higher spin gauge fields described
by completely symmetric tensors. The Lagrangians for these
theories, leading to physical second-order equations, have been
given long ago in \cite{Fronsdal:1978rb}. We show that the
spin-$s$ theory, described in \cite{Fronsdal:1978rb} by a totally
symmetric tensor with $s$ indices and subject to the 
double-tracelessness condition, is dual to a theory with a field
of mixed symmetry type $(D-3, 1, 1, \cdots, 1)$ (one column with
$D-3$ boxes, $s-1$ colums with one box), for which we give
explicitly the Lagrangian and gauge symmetries. This field is also
subject to the double tracelessness condition on any pair of pairs 
of indices. A crucial tool in the analysis is given by the
first-order reformulation of the Fronsdal action due to Vasiliev
\cite{Vasiliev:1980as}, which is in fact our starting point. We
find again that in the self-dual dimension four, the original
description and the dual description are the same.

%**********************************************
\section{Spin-2 duality} 
%**********************************************
\setcounter{equation}{0}
\setcounter{theorem}{0}
\setcounter{lemma}{0}
%****************************
\subsection{Parent actions}
%*****************************
% 
We consider the first-order action \cite{West}
\be \cs [e_{a b}, Y^{ab|}_{~~~
c }] = -2 \int d^Dx \left[Y^{a b|c} \pa_{[a}e_{b]c} -
\frac{1}{2}Y_{ab|c}Y^{ac|b} +\frac{1}{2(D-2)}Y_{ab|}^{~~b}
Y^{ac|}_{~~~c} \right] \label{actioneY} \ee where $e_{ab}$ has
both symmetric and antisymmetric parts and where $ Y^{ab|}_{~~~c}
= - Y^{ba|}_{~~~c}$ is a once-covariant, twice-contravariant mixed
tensor. Neither $e$ nor $Y$ transform in irreducible 
representations of the general linear group since $e_{ab}$ has no
definite symmetry while $Y^{ab|}_{~~~c}$ is subject to no trace
condition. Latin indices run from $0$ to $D-1$ and are lowered or
raised with the flat metric, taken to be of ``mostly plus"
signature $(-,+,\cdots ,+)$. The spacetime dimension $D$ is $\geq
3$. The factor $2$ in front of (\ref{actioneY}) is inserted to
follow the conventions of \cite{Vasiliev:1980as}.

The action (\ref{actioneY}) differs from the standard first-order
action for linearized gravity in which the vielbein $e_{ab}$ and the spin
connection $\omega_{ab \vert c}$ are treated as independent variables by a
mere change of variables $\omega_{ab \vert c} \rightarrow Y^{ab|}_{~~~c}$
such that the coefficient of the antisymmetrized derivative of the
vielbein in the action is just $Y^{ab|}_{~~~c}\,$, up to the inessential 
factor of $-2$. This change of
variables reads 
$$ Y_{ab \vert c} = \omega_{c|a | b} +\h_{ac} \omega^{i}_{~|b| i}-\h_{bc} 
\omega^{i}_{~|a | i}; \;\;
\omega_{a|b | c} = Y_{bc \vert a}+ \frac{2}{D-2}\h_{a[b} 
Y_{c]d \vert }^{~~~d} .$$
It was
considered (for full gravity) previously in \cite{West}.

By examining the equations of motion for $Y^{ab|}_{~~~c}$, one 
sees that $Y^{ab|}_{~~~c}$ is an auxiliary field that can be
eliminated from the action. The resulting action is \be {\cs
[e_{ab}]= 4 \int d^Dx \left[C_{ca |}^{~~~a}C^{cb|}_{~~~b} -
\frac{1}{2}C_{ab|c}C^{ac|b}- \frac{1}{4}C_{ab|c}C^{ab|c}\right] }
\label{actione} \ee where $C_{ab|c} = \pa_{[a}e_{b]c}$. This
action depends only on the symmetric part of $e_{ab}$ (the
Lagrangian depends on the antisymmetric part of $e_{ab}$ only 
through a total derivative) and is a rewriting of the linearised
Einstein action of general relativity (Pauli-Fierz action).

{}From another point of view, $e_{ab}$ can be considered in the
action (\ref{actioneY}) as a Lagrange multiplier for the
constraint $\pa_a Y^{ab|}_{~~~c}$ = 0. This constraint can be
solved explicitely in terms of a new field $Y^{abe|}_{~~~~c} =
Y^{[abe]|}_{~~~~~c}$, $Y^{ab|}_{~~~c} = \pa_e Y^{abe|}_{~~~~c}$. 
The action then becomes \be \cs [Y^{abe|}_{~~~~c} ] = 2 \int d^Dx
\left[\frac{1}{2}Y_{ab|c}Y^{ac|b} -\frac{1}{2(D-2)}Y_{ab|}^{~~b}
Y^{ac|}_{~~~c} \right] \label{actionY} \ee where $Y^{ab|c}$ must
now be viewed as the dependent field $Y^{ab|c} = \pa_e Y^{abe|c}$.
The field $Y^{abe|}_{~~~~c}$ can be decomposed into irreducible
components: $Y^{abe|}_{~~~~~c} = X^{abe|}_{~~~~c} +
\d_c^{[a}Z^{be]}$, with $X^{abc|}_{~~~~c}=0$, $X^{abe|}_{~~~~c} = 
X^{[abe]|}_{~~~~~c}$ and $Z^{be} = Z^{[be]}$. A direct but
somewhat cumbersome computation shows that the resulting action
depends only on the irreducible component $X^{abe|}_{~~~~c} $,
i.e. it is invariant under arbitrary shifts of $Z^{ab}$ (which
appears in the Lagrangian only through a total derivative). One
can then introduce in $D \geq 4$ dimensions the field $T_{a_1
\cdots \,a_{D-3}|c}= \frac{1}{3!} \epsilon_{a_1 \cdots 
\,a_{D-3}efg}X^{efg|}_{~~~~c}$ with $T_{[a_1 \cdots \,a_{D-3}
|c]}= 0$ because of the trace condition on $X^{efg|}_{~~~~c}$, and
rewrite the action in terms of this field\footnote{For $D=3$, the
field $X^{efg|}_{~~~~c}$ is identically zero and the dual
Lagrangian is thus ${\cal L} = 0$. The duality transformation
relates the topological Pauli-Fierz Lagrangian to the topological
Lagrangian ${\cal L} = 0$. We shall assume $D> 3$ from now on.}. 
Explicitly, one finds the action given in
\cite{Curtright,Aulakh:cb}: 
\bqn 
\cs[T_{a_1 \cdots \,a_{D-3}|c} ]
&=&- \frac{1}{(D-3)!} \int d^D x \Big[ \pa^e T^{b_1 ...
b_{D-3}|a}\pa_e T_{b_1 ... b_{D-3}|a} -\pa_e T^{b_1 ... b_{D-3}|e
}\pa^f T_{b_1 ... b_{D-3}|f }
\nonumber \\
&-& (D-3)[ - 3 \pa_e T^{e b_2 ... b_{D-3}|a}\pa^f 
T_{f b_2 ... b_{D-3}|a}-2 T_{g}^{~ b_2 ... b_{D-3}|g }\pa^{e f }
T_{e b_2 ... b_{D-3}|f } \nonumber \\
&-&\pa^e T_{g}^{~ b_2 ... b_{D-3}|g }\pa_e
T^{f}_{~ b_2 ... b_{D-3}|f } +(D-4)\pa_ e T_{g}^{~ e b_3 ... b_{D-3}|g }\pa^h
T^{f}_{~~ h b_3 ... b_{D-3}|f }] \Big]\,.
\label{action0T}
\eqn 
By construction, this dual action is equivalent to the initial
Pauli-Fierz action
for linearised general relativity. We shall compare it in the next 
subsections to
the Pauli-Fierz ($D=4$) and Curtright ($D=5$) actions.

One can notice that the equivalence between the actions 
(\ref{actione}) and (\ref{actionY}) can also be proved using the
following parent action: \be \cs [C_{ab|c},Y_{abc|d} ] = 4 \int
d^Dx \left[ -
\frac{1}{2}C_{ab|c}\pa_dY^{dab|c}+C_{ca|}^{~~~a}C^{cb|}_{~~~b} -
\frac{1}{2}C_{ab|c}C^{ac|b}- \frac{1}{4}C_{ab|c}C^{ab|c}\right],
\label{actionCY}\ee where $C_{ab|c}=C_{[ab]|c}$ and $Y_{abc|d}
=Y_{[abc]|d} $. The field $ Y_{abc|d}$ is then a Lagrange
multiplier for the constraint $\pa_{[a} C_{bc]|d}$ = 0, this
constraint implies $C_{ab|c}=\pa_{[a}e_{b]c}$ and, eliminating it, 
one finds that the action (\ref{actionCY}) becomes the action
(\ref{actione}). On the other hand, $C_{ab|c}$ is an auxiliary
field and can be eliminated from the action (\ref{actionCY}) using
its equation of motion, the resulting action is then the action
(\ref{actionY}).

\subsection{Gauge symmetries}
The gauge invariances of the action (\ref{actione}) are known:
$\d e_{ab} = \pa_a\x_b + \pa_b\x_a +\o_{ab}$, where $\o_{ab}=\o_{[ab]}.$
These transformations can be extended to the auxiliary fields 
(as it is always the case \cite{aux}) leading to the gauge invariances of
the parent action (\ref{actioneY}): \bqn
\d e_{ab} = \pa_a\x_b + \pa_b\x_a, \label{gtdiff1}\\
\d Y^{ab|}_{~~~~d}= - \, 6 \, \pa_c \, \pa^{[a} \x^b \d_d^{c]}
\label{gtdiff2}
\eqn
and
\bqn 
\d e_{ab} = \o_{ab},\\
\d Y^{ab|}_{~~~~d}= 3 \,\pa_c \, \o^{[ab}\d_d^{c]}. \eqn
Similarly, the corresponding invariances for the
other parent action (\ref{actionCY}) are:
\bqn
\d C_{ab|c} = \pa_c \pa_{[a}\x_{b]},\\
\d Y^{abc|}_{~~~~~d}= -\, 6 \,\pa^{[a} \x^b \d_d^{c]} 
\eqn
and
\bqn
\d C_{ab|c} = \pa_{[a}\o_{b]c},\\
\d Y^{abc|}_{~~~~d}= 3 \, \o^{[ab}\d_d^{c]} . \eqn
These transformations affect only the irreducible
component $Z^{be}$ of $Y^{abe|}_{~~~~c}$.
[Note that one can redefine the gauge parameter $\omega_{ab}$ 
in such a way that $\d e_{ab} = \pa_a\x_b + \omega_{ab}$. In that case,
(\ref{gtdiff1}) and (\ref{gtdiff2}) become simply
$\d e_{ab} = \pa_a\x_b$, $\d Y^{ab|}_{~~~~d} = 0$.]

Given $Y^{ab|}_{~~~c}$, the equation $Y^{ab|}_{~~~c} = \pa_e Y^{abe|}_{~~~~c}$ 
does not
entirely determine $Y^{abe|}_{~~~~c}$. Indeed $Y^{ab|}_{~~~c}$ is
invariant under the transformation \be \d Y^{abe|}_{~~~~c}= \pa_f
\,( \phi^{abef|}_{~~~~~c}) \label{invceY}\ee of $Y^{abe|}_{~~~~c}$, with
$\phi^{abef|}_{~~~~~c} = \phi^{[abef]|}_{~~~~~~c}$. As the action
(\ref{actionY}) depends on $Y^{abe|}_{~~~~~c}$ only through
$Y^{ab|}_{~~~~c}$, it is also invariant under the gauge
transformations (\ref{invceY}) of the field $Y^{abe|c}$. 
In addition, it is invariant under arbitrary shifts of the irreducible
component $Z^{ab}$,
\be
\d Y^{abc|}_{~~~~d}= 3 \, \o^{[ab}\d_d^{c]} .\ee
The gauge invariances of the action (\ref{action0T}) involving only
$X^{abe|}_{~~~~c}$ (or equivalently, 
$T_{a_1 \cdots \,a_{D-3}|c}$) are simply
(\ref{invceY}) projected on the irreducible component
$X^{abe|}_{~~~~c}$ (or $T_{a_1 \cdots \,a_{D-3}|c} $).

It is of interest to note that it is the same $\omega$-symmetry that
removes the antisymmetric component of the tetrad in the action
(\ref{actione}) (yielding the Pauli-Fierz action for $e_{(ab)}$)
and the trace $Z^{ab}$ of the field $Y^{abe|}_{~~~~~c}$ (yielding 
the action (\ref{action0T}) for $T_{a_1 \cdots \,a_{D-3}|c} $ (or
$X^{abe|}_{~~~~c}$)). Because it is the same invariance that
is at play, one cannot
eliminate simultaneously both $e_{[ab]}$ and the trace of $Y^{ab|}_{~~~c}$
in the parent actions, even though these fields can
each be eliminated individually in their corresponding ``children" actions 
(see \cite{West:2002jj} in this context).

\subsection{D=4: ``Pauli-Fierz is dual to Pauli-Fierz"}
In $D=4$ spacetime dimensions,
the tensor $T_{a_1 \cdots \,a_{D-3}|c} $ has just two indices and
is symmetric, $T_{ab} = T_{ba}$. A direct computation shows that the action
(\ref{action0T}) becomes then
\be 
\cs[T_{ab}] =\int d^4 x \, [\pa^a T^{bc}\pa_a T_{bc}-
2 \pa_a T^{ab}\pa^c T_{cb}-2 T_a^{~a} \pa^{bc}T_{bc}
-\pa_a T_b^{~b} \pa^a T_c^{~c}]
\ee
which is the Pauli-Fierz action for the symmetric massless
tensor $T_{ab}$. At the same time, the gauge parameters
$\phi^{abef|}_{~~~~~c}$ can be written as $\phi^{abef|}_{~~~~~c} =
\epsilon^{abef} \gamma_c$ and the gauge transformations reduce to
$\delta T_{ab} \sim \partial_a \gamma_b + \partial_b \gamma_a$, as they
should. 
Our dualization procedure possesses thus the distinct feature, in four spacetime
dimensions, of mapping the Pauli-Fierz action on itself.
Note that the electric (respectively, the magnetic) part of the (linearized) Weyl
tensor of the original Pauli-Fierz field $h_{ab} \equiv e_{(ab)}$ is
equal to the magnetic (respectively, minus the electric)
part of the (linearized) Weyl tensor of
the dual Pauli-Fierz $T_{ab}$,
as expected for duality \cite{DesNep,Hull1}.

An alternative, interesting, dualization procedure has been discussed in 
\cite{CasUrr1}.
In that procedure, the dual theory is described by a different action,
which has an additional antisymmetric field, denoted $\omega_{ab}$. This field 
does
enter non trivially the Lagrangian through its divergence
$\partial^a \omega_{ab}$\footnote{In
the Lagrangian (27) of \cite{CasUrr1},
one can actually dualize the 
field $\omega_{ab}$
to a scalar $\Phi$ (i.e., (i) replace $\partial^a \omega_{ab}$ by a vector 
$k_b$ in the
action; (ii) force $k_b = \partial^a \omega_{ab}$ through a Lagrange multiplier
term $\Phi \partial^a k_a$ where $\Phi $ is the Lagrange multiplier; and (iii)
eliminate the auxiliary field $k_a$ through its equations of motion).
A redefinition of the symmetric field $\tilde{h}_{ab}$ of \cite{CasUrr1} 
by a term $\sim \eta_{ab} \Phi$ enables one to absorb the scalar $\Phi$,
yielding the Pauli-Fierz action for the redefined symmetric field.}.

\subsection{D=5: ``Pauli-Fierz is dual to Curtright"}
In $D=5$ spacetime dimensions, the dual field is $T_{ab|c}= \frac{1}{3!}
\ve_{abefg}X^{efg|}_{~~~~c}$ , and has the
symmetries $T_{ab|c }= T_{[ab]|c} $ and $T_{[ab|c]}= 0.$ The action found 
by substituting this field into (\ref{actionY}) reads 
\bqn
&\cs[T_{ab|c }] = \frac{1}{2} \int d^5 x &[\pa^a T^{bc|d}\pa_a T_{bc|d} 
-2 \pa_a T^{ab|c}\pa^d T_{db|c}
-\pa_a T^{bc|a}\pa^d T_{bc|d} \nonumber \\
&&-4 T_a^{~b|a}\pa^{cd}T_{cb|d}-2 \pa_a T_b^{~c|b}\pa^a T^d_{~~c|d}
-2 \pa_a T_b^{~a|b}\pa^c T^d_{~~c|d}]
\eqn
It is the action given by Curtright in
\cite{Curtright} for such an ``exotic" field. 

The gauge symmetries also match, as can be seen by redefining the
gauge parameters as $\psi_{gc}=
-\frac{1}{4!}\epsilon_{abefg}\phi^{abef|}_{~~~~~c}$. The gauge
transformations become 
\be 
\d T_{ab|c }= -2\pa_{[a}S_{b]c}- \frac{1}{3}
[\pa_aA_{bc}+\pa_bA_{ca}-2\pa_cA_{ab}], 
\ee 
where $\psi_{ab}=S_{ab}+A_{ab}, S_{ab}=S_{ba}, A_{ab}=-A_{ba}$. 
These are exactly the gauge transformations of \cite{Curtright}.

It was known from \cite{Hull1} that the equations of motion for a
Pauli-Fierz field were equivalent to the equations of motion for a
Curtright field, i.e., that the two theories were ``pseudo-dual".
We have established here that they are, in fact, dual. The
duality transformation considered here contains the duality
transformation on the curvatures considered in \cite{Hull1}. 
Indeed, when the equations of motion hold, one has $R_{\m \n \a
\b}[h] \propto \varepsilon_{\m \n \r \s \tau} R^{\r \s
\tau}_{\hspace{.5cm} \a \b}[T]$ where $R_{\m \n \a \b}[h]$
(respectively $R_{\r \s \tau \a \b}[T]$) is the linearized
curvature of $h_{ab}\equiv e_{(ab)}$ (respectively, $T_{ab |c}$).

%
%--------------------------------------------------
\section{Vasiliev description of higher spin fields}
\setcounter{equation}{0}
\setcounter{theorem}{0}
\setcounter{lemma}{0}
%--------------------------------------------------

In the discussion of duality for spin-two gauge fields, a crucial
role is played by the first-order action (\ref{actioneY}), in
which both the (linearized) vielbein and the (linearized)
spin-connection (or, rather, a linear combination of it) are
treated as independent variables. This first-order action is
indeed one of the possible parent actions. In order to extend the
analysis to higher-spin massless gauge fields, we need a similar
description of higher-spin theories. Such a first-order 
description has been given in \cite{Vasiliev:1980as}. In this
section, we briefly review this formulation, alternative to the
more familiar second-order approach of \cite{Fronsdal:1978rb}. We
assume $s >1$ and $D>3$.

\subsection{Generalized vielbein and spin connection}
The set of bosonic fields introduced in \cite{Vasiliev:1980as}
consists of a generalized vielbein $e_{\m\vert a_1\ldots a_{s-1}}$ 
and a generalized spin connection $\o_{\m\vert b\vert a_1\ldots
a_{s-1}}$. The vielbein is completely symmetric and traceless in
its last $s-1$ indices. The spin-connection is not only completely
symmetric and traceless in its last $s-1$ indices but also
traceless between its second and one of its last $s-1$ indices.
Moreover, complete symmetrization in all its indices but the first
gives zero. Thus, one has
\bqn 
&e_{\m\vert a_1\ldots 
a_{s-1}}=e_{\m\vert(a_1\ldots a_{s-1})}\,,~~ e_{\m\vert~b\ldots
a_{s-1}}^{~\;\, b}=0\,,&
\nonumber \\
&\o_{\m\vert b\vert a_1\ldots a_{s-1}}=\o_{\m\vert b\vert (a_1\ldots a_{s-1})}
\,,~~ \o_{\m\vert (b\vert a_1\ldots a_{s-1})}=0\,,&
\nonumber \\
&\o_{\m\vert b\vert ~ c\ldots a_{s-1}}^{~~~\,c}=0\,,~~ \o_{\m\vert
~\vert b\ldots a_{s-1}}^{~~b}=0\,.& \label{algcond1} 
\eqn 
The 
first index of both the vielbein and the spin-connection may be
seen as a spacetime form-index, while all the others are regarded
as internal indices. As we work at the linearized level, no
distinction will be made between both kinds of indices and they
will both be labelled either by Greek or by Latin letters, running
from $0,1,\cdots,D-1$.

The action was originally written in \cite{Vasiliev:1980as} in
four dimensions as \be \cs^s[e,\o]=\int d^4x \, \ve^{\m\n\r\s} \,
\ve_{abc\s} \, \o_{\r\vert}^{~\; b\vert a i_1\ldots i_{s-2}}\Big(
\pa_{\m}e_{\n\vert i_1\ldots i_{s-2}}^{\hspace*{36pt}c}
-1/2\o_{\m\vert\n\vert i_1\ldots i_{s-2}}^{\hspace*{43pt}c}\Big).
\label{Vasaction} \ee By expanding out the product of the two
$\epsilon$-symbols, one can rewrite it in a form valid in any 
number of spacetime dimensions, \bqn &&\cs^s[e,\omega]=- 2 \int
d^Dx\Big[ (B_{a_1[\n\vert\m]a_2\ldots a_{s-1}}
-\frac{1}{2(s-1)}B_{\n\m\vert a_1\ldots a_{s-1}})
K^{\m\n\vert a_1\ldots a_{s-1}}+
\nonumber \\
&&\hspace{3.8cm} (2B^{\r}_{~\m\vert a_2\ldots a_{s-1}\r}+
(s-2)B^{\r}_{~a_2\vert a_3 \ldots a_{s-1}\m\r})
K^{\m\n\vert a_2\ldots a_{s-1}}_{\hspace*{45pt}\n}\Big]
\label{VasaB} \eqn where \be B_{\m b\vert a_1\ldots a_{s-1}}\equiv 
2 \o_{[\m \vert b]\vert a_1 \ldots a_{s-1}} \label{Bdefinition}
\ee and where \be K^{\m\n\vert a_1\ldots a_{s-1}} =
\pa^{[\m}e^{\n]\vert a_1\ldots a_{s-1}}
-\frac{1}{4}B^{\m\n\vert a_1\ldots a_{s-1}}. \ee
The field $B_{\m b\vert a_1\ldots a_{s-1}}$ is antisymmetric in 
the first two indices, symmetric in the last $s-1$ internal
indices and traceless in the internal indices, \be B_{\m b\vert
a_1\ldots a_{s-1}} = B_{[\m b]\vert a_1\ldots a_{s-1}}, \; B_{\m
b\vert a_1\ldots a_{s-1}} = B_{\m b\vert (a_1\ldots a_{s-1})}, \;
B_{\m b\vert a_1\ldots a_{s-2}}^{\hspace{40pt} a_{s-2}} = 0 \ee
but is otherwise arbitrary : given $B$ subject to these
conditions, one can always find an $\o$ so that
(\ref{Bdefinition}) holds \cite{Vasiliev:1980as}.

The invariances of the action (\ref{Vasaction}) are \cite{Vasiliev:1980as}
\bqn
\d e_{\m\vert a_1\ldots a_{s-1}} &=& \pa_{\m}\xi_{a_1\ldots a_{s-1}} 
+\a_{\m\vert a_1\ldots a_{s-1}}\,,
\label{Vasinva1}\\
\d \o_{\m\vert b\vert a_1\ldots a_{s-1}}&=&
\pa_{\m}\a_{b\vert a_1\ldots a_{s-1}}+\S_{\m\vert b\vert a_1\ldots a_{s-1}}\,,
\label{Vasinva2}
\eqn
where the parameters $\a_{\m\vert a_1\ldots a_{s-1}}$ and
$\S_{\m\vert b\vert a_1\ldots a_{s-1}}$ 
possess
the following algebraic properties
\bqn
&\a_{\n |(a_1 ... a_{s-1})}=\a_{\n |a_1 ... a_{s-1}}\,,~~~
\a_{(\n |a_1 ... a_{s-1})}=0\,,~~~
\a^{\n}_{~|\n a_2 ... a_{s-1}}=0 \,,~~~
\a_{\n |a_1 ... a_{s-3}b}^{~~~~~~~~~~~~b}=0, \nonumber \\
&\S_{\m\vert b\vert a_1\ldots a_{s-1}}=\S_{(\m\vert b)\vert 
a_1\ldots a_{s-1}}=\S_{\m\vert b\vert (a_1\ldots a_{s-1})}\,,~~~
\S_{\m\vert (b\vert a_1\ldots a_{s-1})}=0\,,\nonumber \\
&\S^b_{~\vert b\vert a_1\ldots a_{s-1}}=0\,,~~~ \S^b_{~\vert
c\vert b a_2\ldots a_{s-1}}=0\,,~~~ \S_{\m\vert b\vert a_1\ldots
a_{s-3}c}^{\hspace*{53pt}c}=0\,. 
\label{algcond2} 
\eqn 
Moreover, the
parameter $\xi$ is traceless and completely symmetric. The 
parameter $\a$ generalizes the Lorentz parameter for gravitation
in the vielbein formalism.

In the Vasiliev formulation, the fields and gauge parameters are
subject to tracelessness conditions contained in (\ref{algcond1})
and (\ref{algcond2}). It would be of interest to investigate
whether these conditions can be dispensed with as in
\cite{FS1,Francia:2002pt}. 

\subsection{Equivalence with the standard second order formulation}
Since the action depends on $\o$ only through $B$, extremizing it with respect
to $\o$ is equivalent to extremizing it with respect to $B$. Thus, we
can view $\cs^s[e,\o]$ as $\cs^s[e,B]$.
In the action $\cs^s[e,B]$, the 
field $B^{\m\n\vert a_1\ldots a_{s-1}}$ is an auxiliary field.
Indeed, the field equations for $B^{\m\n\vert a_1\ldots a_{s-1}}$
enable one to express $B$ in terms of the vielbein and its derivatives as,
\be
B^{\m\n\vert a_1\ldots a_{s-1}}=2 \pa^{[\m}e^{\n]\vert a_1\ldots a_{s-1}}
\label{sol0forB}
\ee
(the field $\o$ is thus fixed up to the pure gauge component related to $\S$.) 
When substituted into (\ref{VasaB}), (\ref{sol0forB}) gives an action 
$S^s[e,B(e)]$
invariant under (\ref{Vasinva1}).

The field $e_{\m\vert a_1\ldots a_{s-1}}$ can be represented by
\bqn e_{\m\vert a_1\ldots a_{s-1}} &=& h_{\m a_1\ldots a_{s-1}}
+\frac{(s-1)(s-2)}{2s}[\h_{\m (a_1}h_{a_2\ldots a_{s-1})}
-\h_{(a_1 a_2} h_{\m a_3\ldots a_{s-1})}]
\nonumber \\ 
&+&\b_{\m\vert a_1\ldots a_{s-1}}\,, \eqn where $h_{\m a_1\ldots
a_{s-1}}$ is completely symmetric, $h_{a_2\ldots a_{s-1}}=
h^{\m}_{\hspace{5pt} \m \ldots a_{s-1}}$ is its trace, and the
component $\b_{\m\vert a_1\ldots a_{s-1}}$ possesses the
symmetries of the parameter $\a$ in (\ref{Vasinva1}) and thus
disappears from $S^s[e,\o(e)]$. Of course, the double trace $h^{\m 
\n }_{\hspace{10pt} \m \n \ldots a_{s-1}}$ of $h_{\m a_1\ldots
a_{s-1}}$ vanishes. The action $S^s[e(h)]$ is nothing but the one
given in \cite{Fronsdal:1978rb} for a completely symmetric and
double-traceless bosonic spin-$s$ gauge field $h_{\m a_1\ldots
a_{s-1}}$.

In the spin-2 case, the Vasiliev fields are $e_{\mu |a}$ and 
$\o_{\nu |b |a}$ with $\omega_{\nu |b |a} = - \o_{\nu |a |b}$. The
$\S$-gauge invariance is absent since the conditions $\S_{\n| b
|a} = - \S_{\n |a |b}$, $\S_{b |c |a} = \S_{c |b |a}$ imply
$\S_{\n |a| b} = 0$. The gauge transformations read \be \d e_{\n
|a} = \pa_ \n \xi_a + \a_{\n |a}, \; \; \d \o_{\n |b |a} = \pa_\n
\a_{b |a} \ee with $\a_{\n |a} = - \a_{a |\n}$. The relation 
between $\o$ and $B$ is invertible and the action (\ref{VasaB}) is
explicitly given by \be S^2[e,B]=- 2\int d^Dx\Big[
(B_{a[\n\vert\m]}
-\frac{1}{2}B_{\n\m\vert a})
(\pa^{[\m}e^{\n]\vert a}
-\frac{1}{4}B^{\m\n\vert a})
+ 2B^{\r}_{~\m\vert \r} 
(\pa^{[\m}e^{\n]\vert }_{\hspace*{10pt}\n}
-\frac{1}{4}B^{\m\n\vert }_{\hspace*{13pt}\n})\Big]
\ee
The coefficient $Y_{\m \n |a}$ of the antisymmetrized derivative
$\pa^{[\m}e^{\n]\vert a}$ of the vielbein is given in terms
of $B$ by 
\be
Y_{\m \n |a} = B_{a[\m|\n]} -\frac{1}{2}B_{\m \n |a} -
2 \h_{a[\m}B_{\n] b|}^{\hspace*{13pt}b}.
\ee
This relation can be inverted to yield $B$ in terms of $Y$,
\be
B_{\m \n |a} = 2 Y_{a [\m| \n]} - 
\frac{2}{D-2}\h_{a[\m}Y_{\n] b|}^{\hspace*{13pt}b}.
\ee 
Re-expressing the action in terms of
$e_{\mu a}$ and $Y_{\m \n a}$ gives the action (\ref{actioneY})
considered previously.
%
%------------------------------
\section{Spin-$3$ duality} 
\setcounter{equation}{0}
\setcounter{theorem}{0}
\setcounter{lemma}{0}
%------------------------------
%
Before dealing with duality in the general spin-$s$ case,
we treat in detail the spin-$3$ case. 

\subsection{Arbitrary dimension $\geq 4$}
Following the spin-$2$ procedure, we first rewrite the action (\ref{VasaB})
in terms of $e_{\n|\r\s}$ and the coefficient $Y_{\m\n|\r\s}$
of the antisymmetrized derivatives of $e_{\n|\r\s}$ in the action.
In terms of $\o_{\m|\n|\r\s}$, this field is given by
\bqn
Y_{\m\n|\r\s}=2 [\o_{\r|[\n|\m]\s}+\o_{\s|[\n|\m]\r}- 
2\o^{\l}_{~~|[\l|\m](\r}\h_{\s)\n}
+ 2\o^{\l}_{~~|[\l|\n](\r}\h_{\s)\m}]
\eqn
or, equivalently
\be
Y_{\m\n|a_1 a_{2}}=B_{a_1 \m|\n a_2}- \frac {1}{4} 
B_{\m\n|a_1 a_{2}}+ 2 \h_{\m a_1} B^{\l}_{~\n|\l a_2 }
+ \h_{\m a_1} B^{\l}_{~a_2|\l \n }
\label{YB3}
\ee
where antisymmetrization in $\m$, $\n$ and symmetrization in
$a_1$, $a_2$ is understood.
The field $Y_{\m\n|\r\s} $fulfills the algebraic relations 
$Y_{\m\n|\r\s} =Y_{[\m\n]|\r\s} =Y_{\m\n|(\r\s)} $ and
$Y_{\m\n|\b}^{~~~~\b}=0$. 

One can invert (\ref{YB3}) to express the field $B_{\m\n|\r\s}$ in
terms of $Y_{\m\n|\r\s}$. One gets \be B_{\m\n|\r\s} =
\frac{4}{3} \Big[Y _{\m\n|\r\s}+2 [Y _{\r[\m|\n]\s}+Y
_{\s[\m|\n]\r}]+
\frac{2}{D-1}[-2\h_{\r\s}Y_{\l[\m|\n]}^{\hspace{20pt}\l}
+Y^{\l}_{\r|\l[\n}\h_{\m]\s}+Y^{\l}_{\s|\l[\n}\h_{\m]\r}]\Big]
\ee When inserted into the action, this yields \bqn 
&&\cs(e_{\m|\n\r},Y_{\m\n|\r\s} )= -2 \int d^Dx \left\{ \right.
Y_{\m\n|\r\s} \pa^{\m} e^{\n|\r\s}
\nonumber \\
&&+ \frac{4}{3}[\frac{1}{4} Y^{\m\n|\r\s} Y_{\m\n|\r\s}
-Y^{\m\n|\r\s} Y_{\r\n|\m\s}
%\frac{1}{D-1} Y^{\m\n|\r}_{~~~~~\r} Y_{\m\n|\l}^{~~~~\l}+
%\frac{2}{D-1} Y^{\m\n|\r}_{~~~~~\r} Y_{\l\m|\n}^{~~~~\l}
+\frac{1}{D-1} Y^{\r\m|\n}_{~~~~~\r} Y_{\l\n|\m}^{~~~~\l}] \left. 
\right\}\,. \eqn

The generalized vielbein $e_{\n|\r\s}$ may again be viewed as a
Lagrange multiplier since it occurs linearly. Its equations of
motion force the constraints \be \pa^{\m}Y_{\m\n|\r\s}=0 
\label{constraintspin3} \ee The solution of this equation is
$Y_{\m\n|\r\s}=\pa^{\l}Y_{\l\m\n|\r\s}$ where
$Y_{\l\m\n|\r\s}=Y_{[\l\m\n]|\r\s}=Y_{\l\m\n|(\r\s)}$ and
$Y_{\l\m\n|\r}^{~~~~~\r}=0$. The action then becomes \bqn
\cs(Y_{\l\m\n|\r\s})= \frac{8}{3} \int d^Dx [-\frac{1}{4}
Y^{\m\n|\r\s} Y_{\m\n|\r\s} +Y^{\m\n|\r\s} Y_{\r\n|\m\s}
%\frac{1}{D-1} Y^{\m\n|\r}_{~~~~\r} Y_{\m\n|\l}^{~~~~\l}\cr +
%\frac{2}{D-1} Y^{\m\n|\r}_{~~~~\r} Y_{\l\m|\n}^{~~~~\l}
-\frac{1}{D-1} Y^{\r\m|\n}_{~~~~~\r} Y_{\l\m|\n}^{~~~~\l}] \,, 
\eqn
where $Y_{\m\n|\r\s}$ must now be viewed as the dependent field
$Y_{\m\n|\r\s}=\pa^{\l}Y_{\l\m\n|\r\s}$ .

One now decomposes the field $Y_{\l\m\n|\r\s}$ into irreducible
components, \be Y^{\l\n\m|}_{~~~~~\r\s} = X^{\l\n\m|}_{~~~~~\r\s} 
+ \d_{(\r}^{[\l}Z^{\m\n]}_{~~~\s)} \label{decomp0} \ee with
$X^{\l\n\m|}_{~~~~~\r\m}=0$, $X^{\l\n\m|}_{~~~~~\r\s}=
X^{[\l\n\m]|}_{~~~~~\r\s}$,
$X^{\l\n\m|}_{~~~~~\r\s}=X^{\l\n\m|}_{~~~~~(\r\s)}$ and $Z^{\m\n}
_{~~~\s}= Z^{[\m\n]}_{~~~\s}$. Since $Z^{\m\n} _{~~~\s}$ is
defined by (\ref{decomp0}) only up to the addition of a term like
$\delta^{[\m}_{\s} k^{\n]}$ with $k^\n$ arbitrary, one may assume 
$Z^{\m\n}_{~~~\n} = 0$. The new feature with respect to spin $2$
is that the field $Z^{\m\n} _{~~~\s}$ is now not entirely pure
gauge. However, that component of $Z^{\m\n} _{~~~\s}$ which is
not pure gauge is entirely determined by
$X^{\l\n\m|}_{~~~~~\r\s}$.

Indeed, the tracelessness condition $Y^{\l\n\m|}_{~~~~~\r\s}
\eta^{\r \s} = 0$ implies \be Z^{[\l \m \vert \n]} = - 
X^{\l\n\m|}_{~~~~~\r\s} \eta^{\r \s} \label{ZenfonctiondeX} \ee
One can further decompose $Z_{\l \m \vert \n} = \Phi_{\l \m \n} +
\frac{4}{3} \Psi_{[\l \vert \m] \n} $ with $\Phi_{\l \m \n} =
\Phi_{[\l \m \n]}= Z_{[\l \m \vert \n]}$ and $\Psi_{\l \vert \m
\n} = \Psi_{\l \vert (\m \n)} = Z_{\l (\m \vert \n)}$. In
addition, $\Psi_{(\l \vert \m \n)} = Z_{(\l \m \vert \n)}= 0$ and
$\Psi_{\l \vert \m \n} \eta^{\m \n} = Z_{\l \m \vert \n} \eta^{\m 
\n} = 0$. {} Furthermore, the $\alpha$-gauge symmetry reads $ \d
Z_{\l \m \vert \n} = \a_{[\l \vert \m] \n}$ i.e, $\d \Phi_{\l \m
\n} = 0$ and $\d \Psi_{\l \vert \m \n} = \frac{3}{4} \a_{\l \vert
\m \n}$. Thus, the $\Psi$-component of $Z$ can be gauged away
while its $\Phi$-component is fixed by $X$. The only remaining
field in the action is $X^{\l\n\m|}_{~~~~~\r\s}$, as in the
spin-$2$ case.

Also as in the spin-$2$ case, there is a redundancy in the solution of
the constraint (\ref{constraintspin3})
for $Y_{\n\a|\b\g}$, leading to the gauge symmetry
(in addition to the $\alpha$-gauge symmetry)
\be
\d Y^{\l\m\n|}_{~~~~~a_1 a_{2}}~=\pa_\r
\psi^{\r\l\m\n|}_{~~~~~~~a_1 a_{2}} 
\label{invaraspin3}
\ee
where $\psi^{\r\l\m\n|}_{~~~~~~~a_1 a_{2}}$ is antisymmetric in
$\r$, $\l$, $\m$, $\n$ and symmetric in $a_1$, $a_2$ and is traceless
on $a_1$, $a_2$, $\psi^{\r\l\m\n|}_{~~~~~~~a_1 a_{2}} \eta^{a_1 a_2} = 0$.
This gives, for $X$,
\be \d X^{\l\m\n|}_{~~~~~a_1 a_{2}}~=\pa_\r
\big( \psi^{\r\l\m\n|}_{~~~~~~~a_1 a_{2}} + \frac{6}{D-1} \d^{[\l}_{(a_1} 
\psi^{\m\n]\r\s|}_{~~~~~~~a_2) \s} \big) \ee

\subsection{$D=5$ and $D=4$}
One can then trade the
field $X$ for a field $T$ obtained by 
dualizing on the indices $\l$, $\m$, $\n$ with the $\epsilon$-symbol.
We shall carry out the computations
only in the case $D=5$ and
$D=4$, since the case of general dimensions will be covered
below for general spins.
Dualising in $D=5$ gives
$X^{\l\n\m|}_{~~~~~\r\s} = \frac{1}{2}\epsilon^{\l\n\m\a\b}T_{\a\b|\r\s} $
and the action becomes:
\bqn 
\cs(T_{\m\n|\r\s})&=& \frac{2}{3} \int d^5x
[-\pa_{\l}T_{\m\n|\r\s}\pa^{\l}T^{\m\n|\r\s}
+2 \pa^{\l}T_{\l\n|\r\s}\pa_{\m}T^{\m\n|\r\s}+2 \pa^{\r}T_{\m\n|\r\s}\pa_{\l}
T^{\m\n|\l\s}
\nonumber \\
&+& 8 T_{\m\n|\r\s}\pa^{\m\r}T_{\l}^{~\n|\l\s}+2T_{\m\n|\r\s} 
\pa^{\r\s}T^{\m\n|\l}_{~~~~~\l}+4 \pa_{\r}T_{\l}^{~~\n|\l\s}\pa^{\r}
T^{\m}_{~~\n|\m\s}
\nonumber \\
&-&4 \pa_{\n}T_{\l}^{~~\n|\l\s}\pa^{\r}
T^{\m}_{~~\r|\m\s}+4 \pa_{\s}T_{\l}^{~~\n|\l\s}\pa^{\r}T_{\r\n|\m}^{~~~~~\m}
+\pa_{\l}T^{\m\n|\r}_{~~~~~\r}\pa^{\l}T_{\m\n|\s}^{~~~~~\s}] 
\eqn
with $T_{\m\n|\r\s}=T_{\m\n|(\r\s)}=T_{[\m\n]|\r\s}$ and $T_{[\m\n|\r]\s}=0$.
The gauge symmetries of the $T$ field following from
(\ref{invaraspin3}) are
\be
\d T_{\m\n|\r\s} =-\pa_{[\m}\varphi_{\n]|\s\r} +
 \frac{3}{4}[\pa_{[\m}\varphi_{\n|\s]\r}+\pa_{[\m}\varphi_{\n|\r]\s}] \,,
\ee
where the gauge parameter $\varphi_{\a|\r\s} \sim
\epsilon_{\a \l\m\n\t} \psi^{\l\m\n\t|}_{\hspace*{23pt}\r\s}$ is such that 
$\varphi_{\a|\r\s} =\varphi_{\a|(\r\s)} $ and 
$\varphi_{\a|\r}^{\hspace*{13pt}\r}=0$.
The parameter $
\varphi_{\a|\r\s} $ can be decomposed into irreducible components:
$\varphi_{\a|\r\s}=\chi_{\a\r\s}+\phi_{\a(\r|\s)}$ where 
$\chi_{\a\r\s}=\varphi_{(\a|\r\s)}$
and $\phi_{\a\r|\s}=\frac{3}{4} \varphi_{[\a|\r]\s}$ . 
The gauge transformation then reads
\be
\d T_{\m\n|\r\s} =\pa_{[\m}\chi_{\n]\r\s}
+\frac{1}{8}[-2\pa_{[\m}\phi_{\n]\r|\s}+3 \phi_{\m\n|(\s,\r)}]\,, 
\ee
and the new gauge parameters are constrained by the condition
$\chi_{\a|\r}^{\hspace*{13pt}\r}+\phi_{\a|\r}^{\hspace*{13pt}\r}=0$.

These are the action and gauge symmetries for the field $T_{\m\n|\r\s}$ dual to
$e_{(\m\n\r)}$ in $D=5$ and coincide with the ones given in 
\cite{Chung:1987mv,Labastida:1987kw,Burdik:2001hj,Bekaert:2003az}.

In four spacetime dimensions, dualization reads 
$T_{\a\r\s } = \epsilon_{\l \m \n \a} X^{\l\m\n|}_{~~~~~\r\s}$.
The field $T_{\a\r\s}$ is totally symmetric because of
$X^{\l\n\m|}_{~~~~~\r\m}=0$.
The action reads 
\bqn
\cs (T_{\m\n\r}) = -\frac{4}{3}\int d^4 x \Big[ \pa_{\l}T_{\m\n\r}
\pa^{\l}T^{\m\n\r}-3\pa^{\m}T_{\m\n\r}\pa_{\l}T^{\l\n\r}-6T_{\l}^{~~\l\m} 
\pa^{\n\r}T_{\m\n\r} \nonumber \\
-3 \pa_{\l}T_{\m}^{~~\m\n}\pa_{\l}T_{\r}^{~~\r\n}-\frac{3}{2}\pa_{\l}
T^{\l\m}_{~~~\m}\pa_{\n}T^{\n\r}_{~~~\r}\Big]
\eqn
The gauge parameter
$\psi^{\r\l\m\n|}_{~~~~~~~a_1 a_{2}}$ can be rewritten as
$\psi^{\r\l\m\n|}_{~~~~~~~a_1 a_{2}} = (-1/2) 
\epsilon^{\r\l\m\n} k_{a_1 a_{2}}$ where $k_{a_1 a_{2}}$ is
symmetric and traceless. The gauge transformations are, in terms
of $T$, $\d T_{\r\s \a} = \pa_\r k_{\s \a} + \pa_\s k_{\a \r} +
\pa_\a k_{\r \s}$. The dualization procedure yields back the
Fronsdal action and gauge symmetries \cite{Fronsdal:1978rb}. Note
also that the gauge-invariant curvatures of the original field
$h_{\m\n\r}\equiv e_{(\m \n \r)}$ and of $T_{\m\n\r}$, which involve now three
derivatives \cite{deWFreed,Damour:vm}, are again related on-shell 
by an $\epsilon$-transformation 
$R_{\a\b\m\n\r\s}[h]\propto\epsilon_{\a\b\a'\b'}$ 
$R^{\a'\b'}_{\hspace{13pt}\m\n\r\s}[T]$, as they should.
%
%--------------------------------
\section{Spin-s duality}
\setcounter{equation}{0}
\setcounter{theorem}{0}
\setcounter{lemma}{0} 
%--------------------------------
%
The method for dualizing the spin-s theory follows exactly the same pattern
as for spins two and three:
\begin{itemize}
\item First, one rewrites the action in terms of $e$ and $Y$ (coefficient 
of the antisymmetrized derivatives of the generalized vielbein in the action);
\item Second, one observes that $e$ is a Lagrange multiplier for a differential
constraint on $Y$, which can be solved explicitly in terms of a new field
$Y$ with one more index;
\item Third, one decomposes this new field into irreducible components;
only one component (denoted $X$) remains in the action; using the
$\epsilon$-symbol, this component can be replaced by the ``dual field"
$T$.
\item Fourth, one derives the gauge invariances of the dual theory 
from the redundancy in the description of the solution of the constraint
in step 2.
\end{itemize}
We now implement these steps explicitly.

\subsection{Trading $B$ for $Y$}
The coefficient of $\pa^{[\n} e^{\m]|a_1 ... a_{s-1}}$ in the
action is given by \bqn Y_{\m\n|a_1 ... a_{s-1}}&=&B_{a_1 \m|\n 
a_2 ... a_{s-1}}- \frac {1}{2(s-1)} B_{\m\n|a_1 ... a_{s-1}}+ 2
\h_{\m a_1} B^{\l}_{~\n|\l a_2 ... a_{s-1}}
\nonumber \\
&+& (s-2) \h_{\m a_1} B^{\l}_{~a_2|\l \n a_3 ... a_{s-1}}\,,
\label{YYBB} \eqn where the r.h.s. of this expression must be
antisymmetrized in $\m,\n$ and symmetrized in the indices $a_i$. 
The field $Y_{\m\n|a_1 ... a_{s-1}}$ is antisymmetric in $\m$ and
$\n$, totally symmetric in its internal indices $a_i$ and
traceless on its internal indices. One can invert (\ref{YYBB}) to
express $B_{\m\n|a_1 ... a_{s-1}}$ in terms of $Y_{\m\n|a_1 ...
a_{s-1}}$. To that end, one first computes the trace of
$Y_{\m\n|a_1 ... a_{s-1}}$. One gets \bqn && Y^\l_{~~\m |\l a_2
\cdots a_{s-1}} = \frac{D + s -4}{2(s-1)} \left(2 B^\l_{~~\m |\l
a_2 \cdots a_{s-1}} + (s-2) B^\l_{~~(a_2 | a_3 \cdots a_{s-1}) \l
\m } \right) \\ 
&& \Leftrightarrow \; B^\l_{~~\m |\l a_2 \cdots a_{s-1}} =
\frac{2(s-1)^2}{s(D+s-4)} \left( Y^\l_{~~\m |\l a_2 \cdots
a_{s-1}} - \left(\frac{s-2}{s-1}\right) Y^\l_{~~(a_2 | a_3 \cdots
a_{s-1}) \l \m } \right) \eqn Using this expression, one can then
easily solve (\ref{YYBB}) for $B_{\m\n|a_1 ... a_{s-1}}$, \bqn 
B_{\m\n|a_1 ... a_{s-1}} &=& 2 \frac{(s-1)}{s}\Big[ (s-2)
Y_{\m\n|a_1\ldots a_{s-1}} - 2 ( s-1) Y_{\m a_1|\n a_2 ...
a_{s-1}}
\nonumber \\
&+&2 \frac {(s-1)}{(D+s-4)}[(s-2)
\h_{a_1 a_2} Y_{\m\r|\n a_3 ... a_{s-1}}^{~~~~~~~~~~~~\r}
\nonumber \\
&-& (s-2)\h_{a_1 \m} Y_{a_2\r|\n a_3 ...
a_{s-1}}^{~~~~~~~~~~~~~\r}+(s-3) \h_{a_1 \m} Y_{\n\r| a_2 ... 
a_{s-1}}^{~~~~~~~~~~~\r} ] \Big] \label{BBYY} \eqn where the
r.h.s. must again be antisymmetrized in $\m,\n$ and symmetrized in
the indices $a_i$. We have checked (\ref{BBYY}) using FORM
(symbolic manipulation program \cite{Jos}).

The action (\ref{VasaB}) now reads \bqn \cs^s &=& - 2 \int d^D x 
\Big[ Y_{\m\n|a_1 ... a_{s-1}} \pa^{[\n} e^{\m]|a_1 ... a_{s-1}} +
\frac{(s-1)^2}{s} \Big[ -Y_{\m\n|a_1 ... a_{s-1}}Y^{\m a_1|\n a_2
... a_{s-1}}
\nonumber \\
&+&\frac {(s-2)}{2(s-1)}Y_{\m\n|a_1 ... a_{s-1}}Y^{\m\n|a_1 ... a_{s-1}} \
+ \frac{1}{(D+s-4)} [ (s-3) Y_{\m\n |a_2 ... a_{s-2}}^{\hspace{45pt}\m}
Y^{\n\r | a_2 ... a_{s-2}}_{\hspace*{48pt}\r} 
\nonumber \\
&-&(s-2)Y_{\m\n |a_2 ... a_{s-2}}^{\hspace{45pt}\m}
Y^{a_2\r | \n a_3 ... a_{s-2}}_{\hspace{56pt}\r}]\Big]\Big]\,.
\eqn
It is invariant under the transformations (\ref{Vasinva1}) and (\ref{Vasinva2})
\bqn
\d e_{\n |a_1 ... a_{s-1}}&=& \pa_{\m}\xi_{a_1\ldots a_{s-1}} +
\a_{\n |a_1 ... a_{s-1}}
\nonumber \\ 
\d Y^{\m\n|} _{~~~~a_1 ... a_{s-1}}&=& 3 \pa_{\l} \d^{[\l}_{(a_1}
\a^{\m|\n]}_{~~~~a_2 ... a_{s-1})}
\eqn
Recall that $ \a_{\n |a_1 ... a_{s-1}}$ satisfies the relations
\be 
\a_{(\n |a_1 ... a_{s-1})}=0\,,~~~
\a^{\n}_{~|\n a_2 ... a_{s-1}}=0 \,,~~~
\a_{\n |a_1 ... a_{s-3}b}^{~~~~~~~~~~~~b}=0 \,.\label{rela}
\ee
while $\xi_{a_1\ldots a_{s-1}}$ is completely symmetric and traceless.
\subsection{Eliminating the constraint}
The field equation for $e^{\m|a_1 ... a_{s-1}} $ is a
constraint for the field $Y$,
\be \pa^{\n}Y_{\n\m|a_1 ... a_{s-1}}=0 
\label{constraintforY}\ee
which implies: \\
\be Y_{\m\n|a_1 ... a_{s-1}}=\pa^{\l}Y_{\l\m\n|a_1 ... a_{s-1}}
\ee where $Y_{\l\m\n|a_1 ... a_{s-1}}=Y_{[\l\m\n]|a_1 ... a_{s-1}} 
=Y_{\l\m\n|(a_1 ... a_{s-1})}$ and $Y^{\l\m\n|a}_{~~~~~~~a a_3 ...
a_{s-1}}=0$ . If one substitutes the solution of the constraint
inside the action, one gets \bqn &&\cs(Y_{\l\m\n|a_1 ... a_{s-1}})
=- 2 \frac{(s-1)^2}{s} \int d^D x \Big[-Y_{\m\n|a_1 ...
a_{s-1}}Y^{\m a_1|\n a_2 ... a_{s-1}}
\nonumber \\
&&+\frac {(s-2)}{2(s-1)}Y_{\m\n|a_1 ... a_{s-1}}Y^{\m\n|a_1 ... a_{s-1}}
+ \frac{1}{(D+s-4)} [ (s-3) Y_{\m\n |a_2 ... a_{s-2}}^{\hspace{45pt}\m}
Y^{\n\r | a_2 ... a_{s-2}}_{\hspace{48pt}\r} 
\nonumber \\
&&-(s-2)Y_{\m\n |a_2 ... a_{s-2}}^{\hspace{45pt}\m}
Y^{a_2\r | \n a_3 ... a_{s-2}}_{\hspace{58pt}\r}]\Big]\,,
\eqn
where $Y_{\m\n|a_1 ... a_{s-1}} \equiv \pa^{\l}Y_{\l\m\n|a_1 ... a_{s-1}} $. 
This action
is invariant under the transformations
\be 
\d Y^{\l\m\n|}_{~~~~~a_1 ... a_{s-1}}~=3 \, \d^{[\l}_{(a_1}
\a^{\m|\n]}_{~~~~a_2 ... a_{s-1})}\,,\label{invara}
\ee
where $\a_{\n |a_1 ... a_{s-1}}$ satisfies the relations (\ref{rela}),
as well as under the transformations
\be
\d Y^{\l\m\n|}_{~~~~~a_1 ... a_{s-1}}~=\pa_\r
\psi^{\r\l\m\n|}_{~~~~~~~a_1 ... a_{s-1}}\,.\label{invarpsi} 
\ee
that follow from the redundancy of the parametrization
of the solution of the constraint
(\ref{constraintforY}).
The gauge parameter $\psi^{\r\l\m\n|}_{~~~~~~~a_1 ... a_{s-1}}$ is subject
to the algebraic conditions
$
\psi^{\r\l\m\n|}_{~~~~~~~a_1 ... a_{s-1}} = 
\psi^{[\r\l\m\n]|}_{~~~~~~~a_1 ... a_{s-1}}=
\psi^{\r\l\m\n|}_{~~~~~~~(a_1 ... a_{s-1})} $ and $
\psi^{\r\l\m\n|}_{~~~~~~~a_1 a_2 ... a_{s-1}} \eta^{a_1 a_2} = 0$.

\subsection{Decomposing $Y_{\l\m\n|a_1 ... a_{s-1}}$ -- Dual action}
The field $Y_{\l\m\n|a_1 ... a_{s-1}}$ can be decomposed into the following
irreducible components
\be Y^{\l\m\n|}_{~~~~~a_1 ... a_{s-1}}= X^{\l\m\n|}_{~~~~~a_1 ... a_{s-1}}+
\d^{[\l}_{(a_1} Z^{\m\n ]|}_{~~~~a_2 ... a_{s-1})} \label{decompy}\ee
where $X^{\l\m\n|}_{~~~~~ \l a_2 ... a_{s-1}}=0$ ,
$Z^{\m\n |}_{~~~~ \m a_3 ... a_{s-1}}=0$.
The condition $Y^{\l\m\n| a}_{~~~~~~~ a a_3 ... a_{s-1}}=0$ implies
\bqn 
Z^{\m\n |a}_{~~~~~ a a_4 ... a_{s-1}}&=&0\,, \label{zz} \\
Z^{[\m\n |\l]}_{~~~~~~~ a_3 ... a_{s-1}}&=& - \frac{(s-1)}{2}
X^{\m\n \l |a}_{~~~~~~~ a a_3 ... a_{s-1}}\,.\label{zx} \eqn The
invariance (\ref{invara}) of the action involves only the field Z
and reads \bqn &\d X^{\l\m\n|}_{~~~~~a_1 ... a_{s-1}}=0 \cr &\d
Z_{\m\n |a_1 ... a_{s-2}}=\a_{[\m|\n] a_1 ... a_{s-2}} 
\label{inva} \eqn Next, one rewrites $Z_{\m\n |a_1 ... a_{s-2}}$
as \be Z_{\m\n |a_1 ... a_{s-2}} = \frac{3(s-2)}{s} \Phi_{\m\n
(a_1 \vert a_2 ... a_{s-2})} + \frac{2(s-1)}{s} \Psi_{[\m \vert
\n] a_1 ... a_{s-2}} \ee with $\Phi_{\m\n a_1 \vert a_2 ...
a_{s-2}} = Z_{[\m\n |a_1] a_2 ... a_{s-2}}$ and $\Psi_{\m \vert \n
a_1 ... a_{s-2}} = Z_{\m(\n |a_1 ... a_{s-2})}$. So the
irreducible component $\Phi_{\m\n a_1 \vert a_2 ... a_{s-2}}$ of
$Z$ can be expressed in terms of $X$ by the relation (\ref{zx}), 
while the other component $\Psi_{\m \vert \n a_1 ... a_{s-2}}$ is
pure-gauge by virtue of the gauge symmetry (\ref{inva}), which
does not affect $\Phi_{\m\n a_1 \vert a_2 ... a_{s-2}}$ and reads
$\d \Psi_{\m \vert \n a_1 ... a_{s-2}} = (1/2) \a_{\m \vert \n a_1
... a_{s-2}}$ (note that $\Psi_{\m \vert \n a_1 ... a_{s-2}}$ is
subject to the same algebraic identities (\ref{rela}) as $\a_{\m
\vert \n a_1 ... a_{s-2}}$). As a result, the only independent
field appearing in $S(Y^{\l\m\n|}_{~~~~~a_1 ... a_{s-1}}) $ is
$X^{\l\m\n|}_{~~~~~a_1 ... a_{s-1}}$. 

Performing the change of variables 
\be 
X^{\l\m\n|}_{~~~~~a_2 ...
a_s} = \frac{1}{(D-3)!} \epsilon^{\l\m\n b_1 ... b_{D-3}} T_{b_1
... b_{D-3}|a_2 ... a_s}\,, 
\label{decompx} \ee 
the action for this field reads 
\bqn 
\cs &=& - \frac{2(s-1)}{s(D-3)!} \int d^D
x \Big[ \pa^e T^{b_1 ... b_{D-3}|a_2 ... a_s}\pa_e T_{b_1 ... 
b_{D-3}|a_2 ... a_s}
\nonumber \\
&&- (D-3)\pa_e T^{e b_2 ... b_{D-3}|a_2 ... a_s}\pa^f
T_{f b_2 ... b_{D-3}|a_2 ... a_s} \nonumber \\
&&+(s-1)[ -\pa_e T^{b_1 ... b_{D-3}|e a_3 ... a_s}\pa^f
T_{b_1 ... b_{D-3}|f a_3 ... a_s} \nonumber \\
&&-2(D-3) T_{g}^{~ b_2 ... b_{D-3}|g a_3 ... a_s}\pa^{e f }
T_{e b_2 ... b_{D-3}|f a_3 ... a_s} \nonumber \\
&&-(s-2)T^{b_1 ... b_{D-3}|c ~~ a_4 ... a_s}_{~~~~~~~~~~~c}\pa^{e f } 
T_{b_1 ... b_{D-3}|e f a_4 ... a_s} \nonumber \\
&& -(D-3)\pa^e T_{g}^{~ b_2 ... b_{D-3}|g a_3 ... a_s}\pa_e
T^{f}_{~ b_2 ... b_{D-3}|f a_3 ... a_s} \nonumber \\
&&-\frac{1}{2}(s-2) \pa^e
T^{ b_1 ... b_{D-3}|c ~~ a_4 ... a_s}_{~~~~~~~~~~~c} \pa_e
T_{ b_1 ... b_{D-3}|d ~~ a_4 ... a_s}^{~~~~~~~~~~~d} 
\nonumber \\
&&+(D-3)(D-4)\pa_ e T_{g}^{~ e b_3 ... b_{D-3}|g a_3 ... a_s}\pa^h
T^{f}_{~~ h b_3 ... b_{D-3}|f a_3 ... a_s} \nonumber \\
&&- (s-2)(D-3) \pa_ e T_{g}^{~ b_2 ... b_{D-3}|g e a_4 ... a_s} \pa^f
T_{f b_2 ... b_{D-3}|c ~~ a_4 ... a_s}^{~~~~~~~~~~~~c} \nonumber \\
&&+\frac{1}{4}(s-2)(D-3) \pa_e
T^{e b_2 ... b_{D-3}|c ~~ a_4 ... a_s}_{~~~~~~~~~~~~c} \pa^f
T_{f b_2 ... b_{D-3}|d ~~ a_4 ... a_s}^{~~~~~~~~~~~~d}
\nonumber \\ 
&&-\frac{1}{4}(s-2)(s-3) \pa_e
T^{ b_1 ... b_{D-3}|c ~~ e a_5 ... a_s}_{~~~~~~~~~~~c} \pa^f
T_{ b_1 ... b_{D-3}|d ~~ f a_5 ... a_s}^{~~~~~~~~~~~d}]\Big]\,. 
\label{actionduale}
\eqn
The field $T_{b_1... b_{D-3}|a_2 ... a_s} $
fulfills the following algebraic properties,
\begin{eqnarray} 
&& T_{b_1 ... b_{D-3} |a_2 ... a_s} =
T_{[b_1 ... b_{D-3}]|a_2 ... a_s} \\
&& T_{b_1 ... b_{D-3} |a_2 ... a_s} =
T_{b_1 ... b_{D-3} |(a_2 ... a_s)} \\
&&T_{[b_1 ... b_{D-3} |a_2] ... a_s} = 0 \\
&& T_{b_1 ... b_{D-3} |a_2 a_3 a_4 a_5... a_s}
\eta^{a_2 a_3} \eta^{a_4 a_5} = 0 \\ 
&& T_{b_1 ... b_{D-3} |a_2 a_3 a_4... a_s} \eta^{b_1 a_2}
\eta^{a_3 a_4}= 0
\end{eqnarray}
the last two relations following from (\ref{zx}) and (\ref{zz}).

Conversely, given a tensor $T_{b_1 ... b_{D-3} |a_2 ... a_s} $ fulfilling the
above algebraic conditions, one may first reconstruct
$X^{\l\m\n|}_{~~~~~a_2 ... a_s}$ such that $X^{\l\m\n|}_{~~~~~a_2 ... a_s} 
= X^{[\l\m\n]|}_{~~~~~a_2 ... a_s}$, $X^{\l\m\n|}_{~~~~~a_2 ... a_s}
= X^{\l\m\n|}_{~~~~~(a_2 ... a_s)}$
and $X^{\l\m\n|}_{~~~~~\n a_3 ... a_s} = 0$. One then gets the
$\Phi$-component of $Z^{\m\n |}_{~~~~a_2 ... a_{s-1}}$
through (\ref{zx}) and finds that it
is traceless thanks to the double tracelessness conditions on
$T_{b_1 ... b_{D-3} |a_2 ... a_s} $. 

The equations of motion for the action (\ref{actionduale}) are
\be
G_{b_1 ... b_{D-3} |a_2 ... a_s}=0\,, \ee
where
\bqn
G_{b_1 ... b_{D-3} |a_2 ... a_s}=F_{b_1 ... b_{D-3} |a_2 ... a_s}
-\frac{(s-1)}{4}\Big[2(D-3) \h_{b_1 a_2} 
F^c_{~b_2 ... b_{D-3} |c a_3 ... a_s}\nonumber \\
+(s-2)\h_{ a_2 a_3}
F^{\hspace{42pt}c}_{b_1 ... b_{D-3} |c ~ ~a_4 ... a_s} \Big], 
\nonumber 
\eqn
and
\bqn
F_{b_1 ... b_{D-3} |a_2 ... a_s} &= &\pa_c \pa^c 
T_{b_1 ... b_{D-3} |a_2 ... a_s} \nonumber \\
&-&(D-3)\pa_{b_1}\pa^c T_{c b_2 ... b_{D-3} |a_2 ... a_s}
-(s-1)\pa_{a_2}\pa^c T_{b_1 ... b_{D-3} |ca_3 ... a_s} \nonumber \\
&+&(s-1) \Big[(D-3)\pa_{a_2 b_1}T^{c}_{~b_2 ... b_{D-3} |c a_3 ... a_s}+ 
\frac{(s-2)}{2}\pa_{a_2 a_3}
T^{\hspace{42pt}c}_{b_1 ... b_{D-3} |c~~a_4 ... a_s}\Big]\,,\nonumber
\eqn
and where the r.h.s. of both expressions has to be antisymmetrised in 
$b_1 ... b_{D-3}$
and symmetrised in $a_2 ... a_s$.

\subsection{Gauge symmetries of dual theory}
As a consequence of (\ref{invarpsi}), (\ref{decompy}) and
(\ref{decompx}), the dual action is invariant under the gauge 
transformations: \bqn \d T_{b_1 ... b_{D-3}|a_2 ... a_s} =
\pa_{[b_1}\phi_{b_2 ... b_{D-3}]|a_2 ... a_s} +
\frac{(s-1)(D-2)}{(D+s-4)}\pa_f \phi_{c_1 ... c_{D-4}|g a_3 ...
a_s} \d^{[fg c_1 ... c_{D-4}]}_{[a_2 b_1 ... b_{D-3}]}\,,
\nonumber \eqn where the r.h.s. must be symmetrized in the
indices $a_i$ and where the gauge parameter $\phi_{b_1 ...
b_{D-4}|a_2 ... a_s} \sim \epsilon_{ b_1 ... b_{D-4} \r \l \m \n}
\psi^{\r\l\m\n \vert }_{~~~~~~a_2 ... a_{s}}$ is such that 
$\phi_{b_1 ... b_{D-4}|a_2 ... a_s} = \phi_{[b_1 ... b_{D-4}]|a_2
... a_s}=\phi_{b_1 ... b_{D-4}|(a_2 ... a_s)}$, and
$\phi^{~~~~~~~~~~a}_{b_1 ... b_{D-4}|~~ a a_4 ... a_s} =0$.

This completes the dualization procedure and provides the dual
description, in terms of the field $T_{b_1 ... b_{D-3}|a_2 ... 
a_s} $, of the spin-s theory in $D$ spacetime dimensions. Note that
in four dimensions, the field $T_{b_1|a_2 ... a_s}$ has $s$
indices, is totally symmetric and is subject to the double
tracelessness condition. One gets back in that case the original
Fronsdal action, equations of motion and gauge symmetries.

%**********************************************************
\section{Comments on interactions} 
\setcounter{equation}{0}
\setcounter{theorem}{0}
\setcounter{lemma}{0}
%**********************************************************

We have investigated so far duality only at the level of the free theories.
It is well known that duality becomes far more tricky in the presence
of interactions. The point is that consistent, local interactions for 
one of the children theories may not be local for the other. For instance,
in the case of $p$-form gauge theories,
Chern-Simons terms are in that class since they
involve ``bare" potentials. An exception where the same interaction is
local on both sides is given by the Freedman-Townsend model
\cite{Free-Town} in four dimensions, where duality relates
a scalar theory (namely, non-linear 
$\sigma$-model) to an interacting $2$-form theory.

It is interesting to analyse the difficulties at the level of the
parent action. We consider the definite case of spin-$2$. The
second-order action $\cs[e_{ab}]$ (Eqn (\ref{actione})) can of
course be consistently deformed, leading to the Einstein action.
One can extend this deformation to the action (\ref{actioneY})
where the auxiliary fields are included (see e.g. \cite{West}). In 
fact, auxiliary fields are never obstructions since they do not
contribute to the local BRST cohomology \cite{aux,BBH}. The
problem is that one cannot go any more to the other single-field
theory action $\cs[Y]$. The interacting parent action has only
one child. The reason why one cannot get rid of the vielbein
field $e_{a \m}$ is that it is no longer a Lagrange multiplier. 
The equations of motion for $e_{a \m}$ are not constraints on $Y$,
which one could solve to get an interacting, local theory on the
$Y$-side (the possibility of doing so is in fact prevented by the
no-go theorem of \cite{Bekaert:2002uh}). Rather, they mix both $e$
and $Y$. Thus, one is prevented from ``going down" to $\cs[Y]$. At
the same time, the other parent action corresponding to
(\ref{actionCY}) does not exist once interactions are switched
on. By contrast, in the Freedman-Townsend model, the Lagrange 
multiplier remains a Lagrange multiplier.

\section{Conclusions}
\setcounter{equation}{0} \setcounter{theorem}{0}
\setcounter{lemma}{0} In this paper, we have analyzed duality for
massless gauge theories with spin $\geq 2$. We have shown how to 
dualize such theories, replacing the original description in
terms of a totally symmetric tensor with $s$ indices by a dual
description involving a tensor with mixed Young symmetry type
characterized by one columns with $D-3$ boxes and $s-1$ columns
with one box. Our results encompass previous analyses where
duality was studied at the level of the curvatures and equations
of motion, but not at the level of the action.

A crucial role is played in the approach by the first-order
formulation due to Vasiliev \cite{Vasiliev:1980as}, which provides
the ``parent action" connecting the two dual formulations (up to
minor redefinitions). First-order formulations associated with
exotic tensor gauge fields were considered recently in
\cite{Zinoviev:2003ix}. As a by-product of the analysis, we
reproduce the known local actions leading to second-order field 
equations for ``exotic" tensor gauge fields transforming in the
representation of the linear group characterized by a Young
diagram with one column with $k$ boxes and $m$ columns with one
box and subject to double-tracelessness conditions.

We have considered here original gauge theories described by
totally symmetric gauge fields only. It would be of interest to
extend the construction to more general tensor gauge fields. This
problem is currently under investigation.

\section*{Acknowledgements}
We are grateful to Peter West for useful discussions.
Work supported in part by 
the ``Actions de Recherche Concert\'ees", a "P\^ole
d'Attraction Interuniversitaire" (Belgium), by IISN-Belgium
(convention 4.4505.86) and by the European Commission RTN
programme HPRN-CT-00131.

\end{document}